# Tail state formation in solar cell materials: First principles analyses of zincblende, chalcopyrite, kesterite and hybrid perovskite crystals

Mitsutoshi Nishiwaki,[1] Keisuke Nagaya,[1] Masato Kato,[1] Shohei Fujimoto,[1] Hitoshi Tampo,[2] Tetsuhiko Miyadera,[2] Masayuki Chikamatsu,[2] Hajime Shibata,[2] and Hiroyuki Fujiwara[1*]

[1]*Department of Electrical, Electronic and Computer Engineering, Gifu University, 1-1 Yanagido, Gifu 501-1193, Japan*

[2]*Research Center for Photovoltaics, National Institute of Advanced Industrial Science and Technology (AIST), Central2, 1-1-1 Umezono, Tsukuba, Ibaraki 305-8568, Japan*

**Abstract**

Tail state formation in solar cell absorbers leads to a detrimental effect on solar cell performance. Nevertheless, the characterization of the band tailing in experimental semiconductor crystals is generally difficult. In this article, to determine the tail state generation in various solar cell materials, we have developed a quite general theoretical scheme in which the experimental Urbach energy is compared with the absorption edge energy derived from density functional theory (DFT) calculation. For this purpose, the absorption spectra of solar cell materials, including CdTe, $CuInSe_2$ (CISe), $CuGaSe_2$ (CGSe), $Cu_2ZnSnSe_4$ (CZTSe), $Cu_2ZnSnS_4$ (CZTS) and hybrid perovskites, have been calculated by DFT particularly using very-high-density k meshes. As a result, we find that the tail state formation is negligible in CdTe, CISe, CGSe and hybrid perovskite polycrystals. However, coevaporated CZTSe and CZTS layers exhibit very large Urbach energies, which are far larger than the theoretical counterparts. Based on DFT analysis results, we conclude that the quite large tail state formation observed in the CZTSe and CZTS originates from extensive cation disordering. In particular, even a slight cation substitution is found to generate unusual band fluctuation in CZTS(Se). In contrast, $CH_3NH_3PbI_3$ hybrid perovskite shows the sharpest absorption edge theoretically, which agrees with experiment.

## I. INTRODUCTION

Potential fluctuation in the band edge region of solar cell absorbers is quite detrimental for the performance of photovoltaic devices, reducing their open-circuit voltages ($V_{oc}$) rather significantly [1-4]. In particular, the creation of the tail states generally leads to serious increase of $V_{oc}$ loss defined by $V_{loss}=E_g/e-V_{oc}$, where $E_g$ shows the band gap of the absorber material. In fact, for hydrogenated amorphous silicon (a-Si:H), a quite large $V_{loss}$ of ~0.8 V is observed due to the extensive tail-state generation induced by the random network [5]. In contrast, high efficiency GaAs and Cu(In,Ga)Se$_2$ solar cells exhibit small $V_{loss}$ in a range of 0.3~0.37 V [6,7], in part due to the lower tail state formation.

The generation of the tail state can be characterized quantitatively through the evaluation of the absorption tail generally expressed by $\alpha \propto \exp(E/E_0)$, where $\alpha$ is the absorption coefficient and $E_0$ shows the slope of the absorption tail. Specifically, $E_0$ obtained experimentally defines the Urbach energy ($E_0=E_U$) [8]. It has already been reported that $E_U$ shows a direct correlation with $V_{loss}$ and the smaller $E_U$ (i.e., sharper absorption edge) is favorable to suppress $V_{loss}$ [2]. Nevertheless, since $E_U$ is determined as a slope of experimental absorption spectra, $E_U$ includes the contributions of the density of states (DOS) derived from (i) valence and conduction bands and (ii) non-ideal tail states of absorber materials. Unfortunately, the separation of these contributions has been rather difficult and detailed $E_U$ analyses of solar cell materials have not been performed yet.

For the detailed interpretation of material optical properties, on the other hand, density functional theory (DFT) calculation has been performed widely [9-15]. However, such DFT analyses have so far been employed to determine $E_g$ [9,10] and overall optical transitions in materials [9,11-15] and only a very few DFT studies have focused on the characterization of band tailing in experimental materials [16]. This is primarily because of the requirement of the complex analysis process and the extensive DFT calculation cost. Moreover, to deduce the band-edge absorption accurately in DFT, the computer calculation using high density k points is generally necessary.

In this study, in an attempt to characterize the tail state formation and the resulting $V_{oc}$ loss in photovoltaic devices, we have developed a general theoretical approach based on very-high-density-k-mesh DFT calculations performed for various absorber materials, including binary zincblende (CdTe), ternary chalcopyrite [CuInSe$_2$ (CISe) and CuGaSe$_2$ (CGSe)] and quaternary kesterite [Cu$_2$ZnSnSe$_4$ (CZTSe) and Cu$_2$ZnSnS$_4$ (CZTS)] and hybrid perovskite [CH$_3$NH$_3$PbI$_3$ (MAPbI$_3$) and HC(NH$_2$)$_2$PbI$_3$ (FAPbI$_3$)] materials. By applying DFT for the absorption-spectrum calculation, DFT-derived tail absorption energy ($E_0=E_{DFT}$) has been determined. In particular, we have evaluated the $E_U/E_{DFT}$ ratio, from which the potential fluctuation of experimental materials is determined systematically. For CdTe, CISe, CGSe and hybrid perovskites, the experimental absorption edges are reproduced quite well by DFT (i.e., $E_U \sim E_{DFT}$) and MAPbI$_3$ shows the sharpest band edge theoretically. In contrast, experimental CZTSe and CZTS crystals exhibit very large tail absorption ($E_U >> E_{DFT}$), which cannot be explained by single phase formation of kesterite. Based on DFT analysis results, we have attributed the extraordinary large potential fluctuations observed in the CZTSe and CZTS to the tail state formation near the conduction band by Cu-Zn cation disordering.

## II. DFT ANALYSIS

### A. Analysis of tail states by DFT

Figure 1 schematically shows the tail-state analysis using DFT. As known well [8], the variation of $\alpha$ near the band edge region can be described by

$$\alpha(E) = \alpha_0 \exp(E/E_0). \qquad (1)$$

The $\alpha_{Ex}$ and $\alpha_{DFT}$ in Fig. 1 represent $\alpha$ obtained from experiment and DFT calculation, respectively. In general, $\alpha_{Ex}$ shows finite values at $E<E_g$ due to the tail-state formation and the tail energy obtained in this region defines $E_U$ [i.e., $E_0=E_U$ when $\alpha(E)=\alpha_{Ex}(E)$].

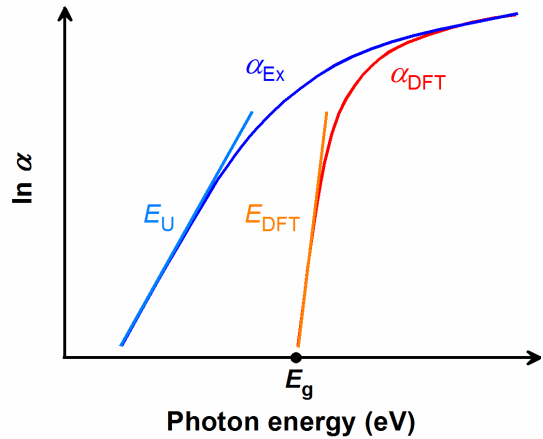

FIG. 1. Schematic representation of absorption-coefficient ($\alpha$) spectra: experimental $\alpha$ ($\alpha_{Ex}$) and DFT $\alpha$ ($\alpha_{DFT}$) spectra. The $E_U$ and $E_{DFT}$ indicate the Urbach energy and DFT-derived absorption edge energy, respectively.

On the other hand, $\alpha_{DFT}$ becomes completely zero at $E<E_g$ and the DFT-derived absorption edge energy can be characterized as $E_0=E_{DFT}$ assuming $\alpha(E)=\alpha_{DFT}(E)$. It should be emphasized that $E_{DFT}$ represents the absorption edge that originates completely from DOS of conduction and valence bands and does not include the contribution of defect states. Thus, when the tail state formation is negligible, we observe $E_U \sim E_{DFT}$, while the extensive tail state generation in experimental crystals leads to $E_U > E_{DFT}$. As a result, the evaluation of the $E_U/E_{DFT}$ ratio enables us to separate the contribution of the conduction/valence band states from that of the defect-derived tail states, allowing the characterization of the tail state formation theoretically.

**B. DFT calculation**

The DFT calculations were performed using the Vienna *ab initio* simulation package (VASP) [17] as well as Advance/PHASE package. For the calculations of the local density approximation (LDA) and generalized gradient approximation within the Perdew-Burke-Ernzerhof scheme (PBE), the Advance/PHASE software was employed, while the VASP software was applied for hybrid functional calculations (PBE0, HSE03 and HSE06) and PBE calculation incorporating spin-orbit coupling (SOC) interaction.

In PBE calculations without SOC, a plane-wave cutoff energy of 350 eV was used, and the structural optimization was made for all absorber crystals until the atomic configuration converged to within 5 meV/Å, except for $FAPbI_3$ (10 meV/Å). The hybrid functional calculations were carried out using a similar condition. For the DFT calculation of CdTe, a two-atom primitive cell was used, while eight-atom primitive cells were employed for the calculations of CISe, CGSe, CZTSe and CZTS. The DFT analyses of $MAPbI_3$ [14] and $FAPbI_3$ [15] have been described in our earlier studies.

The $\varepsilon_2$ spectra are calculated according to

$$\varepsilon_2 = \frac{\hbar e^2}{8\pi^2 \varepsilon_0 m\omega} \int f\delta(E_{c,\mathbf{k}} - E_{v,\mathbf{k}} - \hbar\omega)d\mathbf{k}, \quad (2)$$

where $E_c$ and $E_v$ show the conduction and valence band energies. In the above equation, $f$ represents the oscillator strength defined by

$$f = \frac{2m\omega}{\hbar}\left|\langle\Psi_c|\mathbf{u}\cdot\mathbf{r}|\Psi_v\rangle\right|^2, \quad (3)$$

where $m$ and $\omega$ are electron mass and angular frequency of incident light, respectively. In Eq. (3), $|\Psi_c\rangle$ and $|\Psi_v\rangle$ are the conduction and valence states, whereas $\mathbf{u}$ and $\mathbf{r}$ represent the polarization vector and position operator, respectively. From the calculated

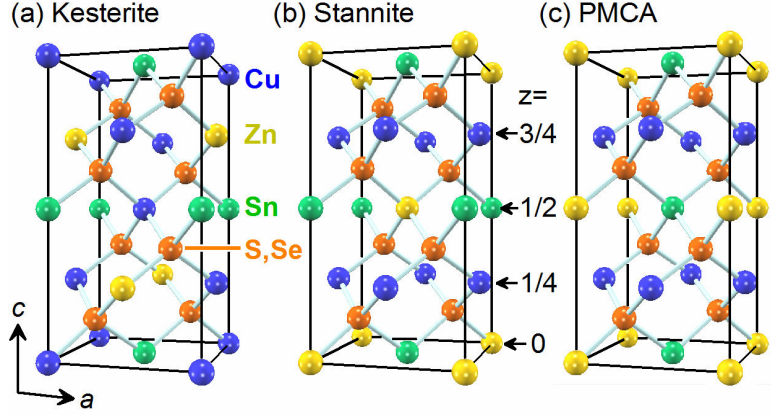

FIG. 2. Three different crystal structures of CZTS(Se) employed in the tail state analyses: (a) kesterite, (b) stannite and (c) primitive-mixed CuAu (PMCA) structures. The arrows indicate the $a$ and $c$ axes of the crystals and z indicates the position of the cationic plane.

$\varepsilon_2$ spectrum, the $\varepsilon_1$ spectrum is obtained using the Kramers-Kronig relations [8]. By employing these $\varepsilon_1$ and $\varepsilon_2$ spectra, the DFT spectra for refractive index $n_{DFT}$ and extinction coefficient $k_{DFT}$ are further calculated, from which $\alpha_{DFT}$ is finally obtained as $\alpha_{DFT}=4\pi k_{DFT}/\lambda$.

In the DFT analyses of CZTS(Se), three different crystal structures shown in Fig. 2 were assumed: i.e., (a) kesterite, (b) stannite and (c) primitive-mixed CuAu (PMCA) structures [18,19]. As confirmed from Fig. 2, CZTS (CZTSe) kesterite crystals are composed of the alternating atomic planes of (Cu,Sn) and (Cu,Zn) with the S (Se) atomic plane in between. The crystal structures of the stannite and PMCA phases are slightly different, and the atomic planes of the cations are separated into the Cu and (Zn,Sn) planes. In the case of PMCA, the (Zn,Sn) planes have a same ordering along the $c$ axis.

### III. RESULTS

**A. Analysis of Urbach energies**

Figure 3 shows experimental $\alpha$ spectra of various solar cell absorbers (open circles) and the result of the Urbach analyses using Eq. (1) (solid lines). For the experimental results, those reported for CdTe [20], CISe [21], CGSe [21], CZTSe [13], CZTS [22], MAPbI$_3$ [14], and FAPbI$_3$ [15] are shown. The optical data of all the materials have been obtained from the thin layers fabricated by coevaporation, except for the CdTe. The $E_U$ of each semiconductor was estimated using a fixed $\alpha$ range of 600-4000 cm$^{-1}$. The maximum range of 4000 cm$^{-1}$ was chosen so that the analyzed region is below $E_g$, whereas the minimum range (600 cm$^{-1}$) corresponds to the sensitivity limit of ellipsometry technique used for the material characterizations [23].

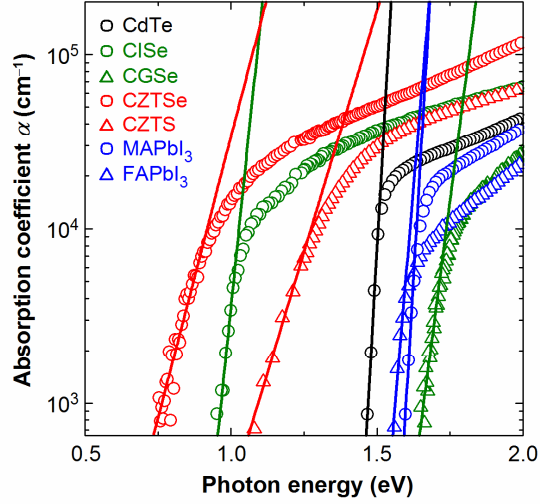

FIG. 3. Experimental $\alpha$ spectra of various solar cell absorbers (open circles) and the result of the Urbach analyses performed assuming $\ln\alpha \propto E/E_U$ (solid lines). For the experimental results, those reported for CdTe [20], CISe [21], CGSe [21], CZTSe [13], CZTS [22], MAPbI$_3$ [14], and FAPbI$_3$ [15] are shown.

All the absorbers in Fig. 3 show similar $\alpha$ values of $10^4$ cm$^{-1}$ near the $E_g$ region. However, the CZTSe and CZTS exhibit strong tail absorption with the broad Urbach slopes, while the other materials show much sharper absorption edges, indicating the distinct tail state formation in the Cu-Zn-Sn-containing quaternary compounds. The quite broad tail absorption in CZT(S)Se has also been confirmed in earlier studies [1,24,25].

**B. DFT calculation of absorption spectra**

In estimating accurate $\alpha$ spectra by DFT, it is essential to employ a very high density k-point mesh as described below. When high-density-k-mesh calculation is performed using hybrid functionals, however, an extensive computing resource is necessary due to the quite high calculation cost of these methods. To constrain the calculation time within the manageable time scale, we calculated $\alpha_{DFT}$ by applying PBE. Nevertheless, the DFT calculation within PBE underestimates $E_g$ severely [10] and thus the underestimated $E_g$ needs to be corrected by blue shifting PBE spectrum. To confirm the validity of this approach, the band structures and optical spectra were calculated using HSE06 and PBE.

Figure 4(a) shows the band structures of CdTe calculated by HSE06 and PBE. In this figure, all the positions of the PBE conduction bands were shifted upward by $\Delta E_g$ so that $E_g$ becomes consistent with that obtained from HSE06. When the value of $\Delta E_g$=0.7 eV is assumed, the shifted PBE conduction bands show excellent agreement with the conduction bands of HSE06.

In Fig. 4(b), the dielectric functions ($\varepsilon_2$ spectra) of CdTe calculated by HSE06 and PBE are compared with the experimental spectrum of Ref. [20]. In these calculations, a 8 × 8 × 8 k mesh (HSE06) and 30 × 30 × 30 k mesh (PBE) were used and the $\varepsilon_2$ spectra obtained from the calculations are shifted so that the onsets of the $\varepsilon_2$ spectra (i.e., $E_0$ transition) match with that of the experimental spectrum. It can be seen that the HSE06 and PBE spectra reproduce the overall experimental spectrum quite well. In particular, in the energy region between the $E_0$ ($E_g$) and $E_1$ transitions (1.5≤$E$≤3.3 eV), both DFT spectra are almost identical. In these DFT calculations, the small transition peak at $E$=3.9 eV, observed experimentally, is not present, but this peak can be reproduced by incorporating the SOC interaction (see Supplementary Fig. 1). The result of Fig. 4(b) indicates that the optical spectrum in the band-edge transition region can be reproduced well by a shifted PBE spectrum.

To find the effect of the k point mesh density on $\alpha_{DFT}$, we further calculated $f$ using Eq. (3). Figure 4(c) shows normalized $f$ in the zincblende Brillouin zone of CdTe obtained using PBE. In particular, $f$ of Fig. 4(c) was calculated for the transition from the first valence band to the first conduction band, which characterizes the band edge absorption (see Supplementary Fig. 2). The numerical values for the contours (black lines) indicate the energy separation between the first valence and conduction bands, and its energy separation is consistent with the optical spectrum of Fig. 4(b). The $E_0$ (1.5 eV), $E_1$ (3.3 eV) and $E_2$ (5.1 eV) transitions in Fig. 4(b) correspond to the transitions at the $\Gamma$, L and X points, respectively [26]. As confirmed from Fig. 4(c), the light absorption near $E_g$ is highly localized near the $\Gamma$ point, indicating that very high density k-point mesh is necessary for the accurate estimation of $\alpha_{DFT}$ and $E_{DFT}$.

In fact, when $\alpha_{DFT}$ was calculated by PBE using different k point meshes [open circles in Fig. 4(d)], the absorption edge of $\alpha_{DFT}$ becomes sharper with increasing k mesh density, and the $E_{DFT}$ analysis (solid lines) also shows the reduction of $E_{DFT}$. In Fig. 4(e), the variations of $E_{DFT}$ with the number of total k points in the PBE calculation are summarized. For the calculation of the two-atom primitive cell (CdTe), $E_{DFT}$ shows a saturation trend at k>$10^4$, whereas for eight-atom primitive cells (CISe and CZTS) and hybrid perovskites we observe the convergence at a slightly lower k of 4×$10^3$.

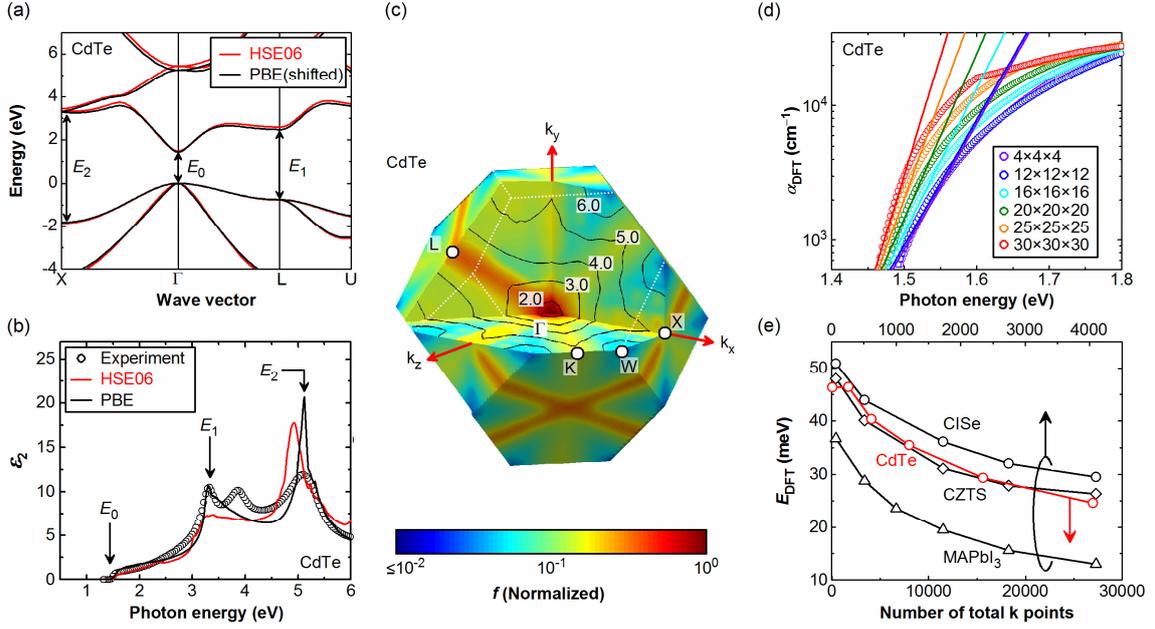

FIG. 4. (a) Band structures of CdTe calculated by HSE06 and PBE, (b) $\varepsilon_2$ spectra of CdTe calculated by HSE06 and PBE, (c) normalized oscillator strength (*f*) of CdTe in the zincblende Brillouin zone, (d) variation of $\alpha_{DFT}$ with k point mesh used in the DFT calculation, and (e) variation of $E_{DFT}$ with the number of total k points used in the DFT calculation. In (a), all the conduction bands of the PBE result are shifted upward by $\Delta E_g$=0.7 eV. In (c), the result corresponds to *f* for the optical transition from the first valence band to the first conduction band. The black lines indicate the cross-section of the contour for the energy separation between the first valence and conduction bands. In (d), the solid lines indicate the result of $E_{DFT}$ analysis.

Unfortunately, the necessity of high k-point mesh density for the precise determination of $E_{DFT}$ is very disadvantageous for HSE06 due to the quite high computational cost. In this study, therefore, we estimated $\alpha_{DFT}$ and $E_{DFT}$ from shifted PBE spectra. In actual PBE calculations, we employed a 30 × 30 × 30 k mesh for CdTe and a 16 × 16 × 16 k mesh for the other solar cell materials, which are the maximum densities allowed in our calculation software.

To validate our approach based on high-mesh-density PBE calculations, systematic DFT calculations were further performed for CdTe using different functionals (i.e., LDA, PBE0, HSE03 and HSE06). We find that the band structures and band-edge DOS are essentially independent of the functional, although the energy position changes depending on the functional [Supplementary Fig. 3(a), 3(b)]. Moreover, $E_{\text{DFT}}$ shows a constant value when the screening parameter of the hybrid functional [9] is changed in a range of $\omega$=0–0.5 Å$^{-1}$ even though $E_g$ varies notably [Supplementary Fig. 3(c)]. In the calculation of HSE06, for the increase of the mixing parameter $a$ [9], the $\alpha$ spectrum shifts toward higher energy but the $E_{\text{DFT}}$ value is quite independent of $a$ [see Supplementary Fig. 3(d)]. As a result, we have confirmed that the band-edge properties and $E_{\text{DFT}}$ are not influenced by the type of DFT functional and the parameter values of the hybrid functional when the same $k$-mesh density is applied for the calculation.

## C. Analysis of band tailing

Figure 5 summarizes the experimental and DFT $\alpha$ spectra of (a) CdTe, (b) CISe, (c) CZTSe and (d) MAPbI$_3$, together with (e) enlarged $\alpha$ spectra near the $E_g$ regions. In this figure, the energy shift values of the PBE spectra (i.e, $\Delta E_g$) are also indicated. As shown in Fig. 5, when a high-density k mesh is employed, $\alpha_{\text{DFT}}$ shows excellent agreement with $\alpha_{\text{Ex}}$ and the overall absorption features are reproduced quite well. When the pure kesterite phase is assumed for the CZTSe, however, the agreement near the band-edge transition region is quite poor and the experimental crystal shows exceptionally large tail absorption, as indicated by the blue color region in Fig. 5(e).

For MAPbI$_3$, on the other hand, the PBE spectra calculated with and without SOC are indicated. As reported earlier [27,28], the SOC interaction alters the MAPbI$_3$ band structure significantly. In particular, when SOC is considered, the band edge position shifts toward lower energy [Supplementary Figs. 4(a) and 4(b)]. When this $E_g$ shift is corrected, however, the $\alpha$ spectrum is quite similar to that obtained without incorporating SOC.

It should be noted that, in the SOC calculation of Fig. 5(d), a smaller k mesh density (8 × 8 × 8 $k$) was used due to the higher computational cost of the SOC calculation, while the calculation without SOC was implemented with a 16 × 16 × 16 $k$, as mentioned above. Thus, the band-edge absorption is slightly broader when the SOC interaction is considered. We find, however, that the $E_{DFT}$ values deduced with and without SOC are essentially the same when the results obtained using the same $k$-mesh density are compared [Supplementary Fig. 4(c)]. Accordingly the influence of the SOC interaction on $E_{DFT}$ is confirmed to be quite minor.

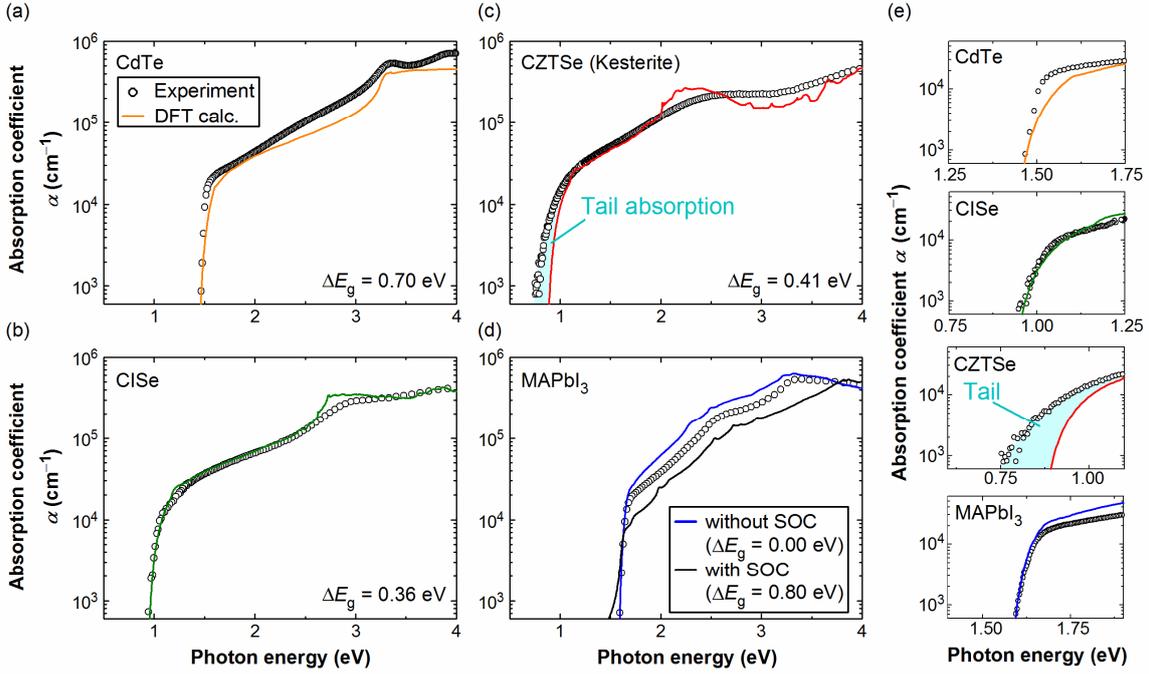

FIG. 5. $\alpha_{Ex}$ (open circles) and $\alpha_{DFT}$ (solid lines) of (a) CdTe, (b) CISe, (c) CZTSe and (d) MAPbI$_3$, together with (e) enlarged $\alpha$ spectra near the $E_g$ regions. The $\alpha_{DFT}$ has been shifted toward higher energy by the energy indicated as $\Delta E_g$ to improve matching with the experimental results. For the DFT calculation of CZTSe, the kesterite structure was assumed. For MAPbI$_3$, the PBE spectrum obtained by incorporating the SOC interaction is also shown.

We further analyzed $E_{DFT}$ from $\alpha_{DFT}$ using the $\alpha$ range employed to estimate $E_U$ (i.e., $\alpha$=600–4000 cm$^{-1}$). Figure 6 compares $E_{DFT}$ obtained from this procedure with $E_U$ estimated in the analyses of Fig. 3 and all the numerical values of Fig. 6 are summarized in Table I. In Fig. 6, a good relationship between $E_U$ and $E_{DFT}$ is confirmed for the zincblende (CdTe), chalcopyrite (CISe, CGSe) and hybrid perovskite polycrystals, indicating the formation of ideal band edges in these materials. In particular, the result of $E_U \sim E_{DFT}$ is quite surprising as the DFT results are obtained assuming perfect crystal structures with zero phonon interaction (i.e., 0 K).

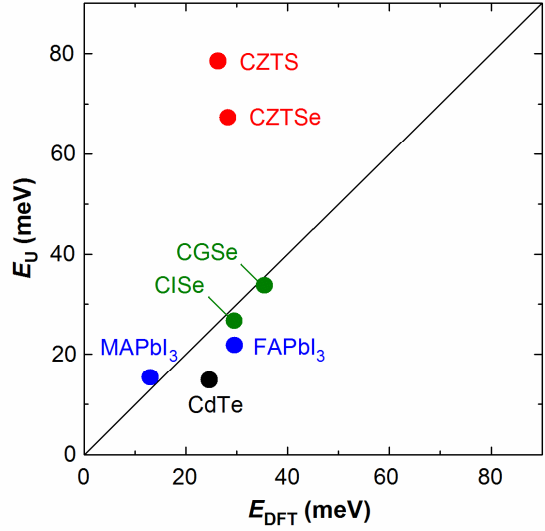

FIG. 6. Urbach energy ($E_U$) as a function of the absorption edge energy derived from DFT ($E_{DFT}$). The $E_U$ values were estimated from the analyses of Fig. 3, whereas $E_{DFT}$ was calculated from $\alpha_{DFT}$ of Fig. 5. For CZTSe and CZTS, the kesterite phases are assumed.

TABLE I. $E_U$ and $E_{DFT}$ values of solar cell materials.

| Materials | $E_U$ (meV) | $E_{DFT}$ (meV) |
|---|---|---|
| CdTe | 14.9 | 24.6 |
| CISe | 26.7 | 29.5 |
| CGSe | 33.8 | 35.4 |
| CZTS | 78.5 | 26.4 (Kesterite) |
|  |  | 27.3 (Stannite) |
|  |  | 32.6 (PMCA) |
| CZTSe | 67.3 | 28.3 (Kesterite) |
|  |  | 33.4 (Stannite) |
| MAPbI$_3$ | 15.4 | 13.0 |
| FAPbI$_3$ | 21.8 | 29.6 |

Among all the absorbers investigated here, MAPbI$_3$ hybrid perovskite shows the sharpest absorption edge theoretically. The quite small $E_{DFT}$ of the hybrid perovskite, compared with the other materials, can be interpreted by the very sharp DOS distribution near the valence and conduction band edges (see Fig. 7). Moreover, $E_{DFT}$ of MAPbI$_3$ is consistent with the experimental value of 15 meV, confirming quite suppressed tail state formation in the experimental perovskite crystal [2]. In general, $E_U$ is quite sensitive to the potential fluctuation generated by i) local band gap variation, ii) defects formed near the valence and conduction band edges, or iii) displacement of atoms at polycrystalline grain boundaries. The quite small $E_U$ of 15 meV, which agrees with the theoretical calculation, indicates clearly that none of the above tail-broadening factors are significant in experimental MAPbI$_3$ polycrystals.

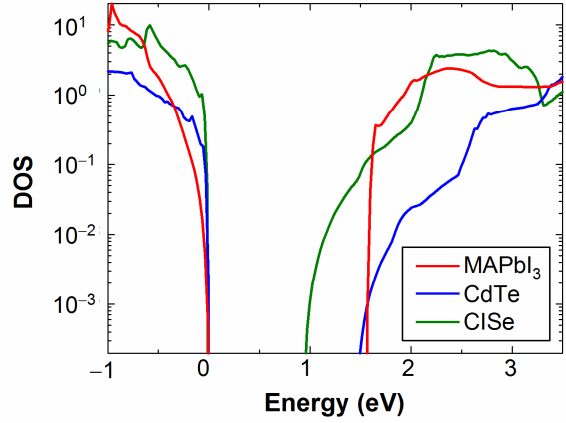

FIG. 7. DOS distribution of the valence and conduction states in MAPbI$_3$, CdTe and CISe.

In Fig. 6, on the other hand, the results for the CZTSe and CZTS are obtained assuming the pure kesterite phases. As confirmed from the results, the $E_U$ values of the CZTSe and CZTS are far larger than $E_{DFT}$, indicating the exceptionally large tail state formation in these crystals. It should be emphasized that $E_{DFT}$ of the kesterite crystal is quite similar to those of stannite and PMCA crystals (see Table I). Accordingly, the quite large $E_U$ values observed experimentally cannot be explained by the formation of a single phase material.

It should be emphasized that $E_U$ of 70–80 meV observed in CZTSe and CZTS is exceptionally large, as even a-Si:H having a complete random structure shows smaller $E_U$ of ~50 meV [5]. In this study, the exceptionally large $E_U$ of the CZTSe and CZTS is attributed to the cation substitution (i.e., Cu-Zn mixing) in the experimental crystals. In particular, we performed first quantitative analysis for the tail absorption observed experimentally in CZTS(Se) based on the DFT calculations.

To validate our hypothesis that the quite extensive tail state formation in the CZTSe and CZTS is induced by the cation disordering, the $\alpha_{DFT}$ spectra of the stannite and PMCA phases are obtained and the $\alpha$ spectrum for a kesterite-stannite-PMCA mixed phase is calculated as a simple weighted average of the three $\alpha_{DFT}$ spectra obtained from the kesterite, stannite and PMCA crystals by neglecting the interaction among the three phases. As shown in Figs. 2(a) and 2(b), the configurations of the Sn and S(Se) atoms in the kesterite and stannite crystals are identical and the Cu-Zn ordering distinguishes these phases. On the other hand, the crystal structure of PMCA is quite close to that of the stannite, as mentioned earlier.

Figure 8(a) shows $\alpha_{Ex}$ (open circles) and the simulated $\alpha$ spectrum (red line) of CZTS. In this figure, the $\alpha_{DFT}$ spectra of the kesterite, stannite and PMCA phases, obtained using the same energy shift value of $\Delta E_g$=0.40 eV, are also shown. As reported previously [18,19,29], the stannite and PMCA crystals have slightly smaller $E_g$, compared with the kesterite crystal, and the whole $\alpha$ spectra of the stannite and PMCA are red shifted by 0.13 eV and 0.31 eV in Fig. 8(a), respectively. Accordingly, the formation of a mixed-phase crystal leads to the large tail absorption by the overlap of the three $\alpha$ spectra. The red line in Fig. 8(a) represents the simulation result obtained from the fitting analysis assuming the three phase mixture and, when we assume the volume ratio of kesterite:stannite:PMCA= 0.35±0.10:0.45±0.05:0.20±0.05, the calculated result shows excellent agreement with the experimental result.

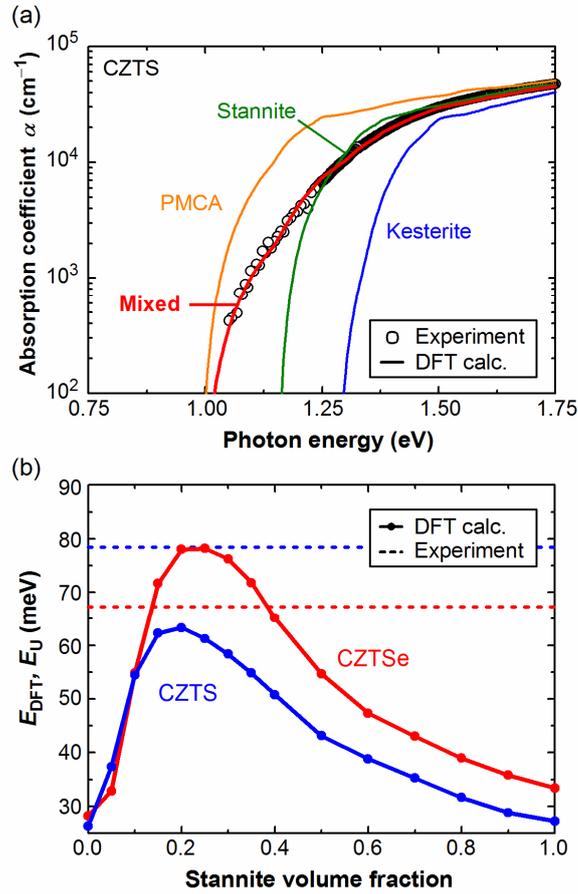

FIG. 8. (a) $\alpha_{DFT}$ spectra of CZTS obtained assuming the kesterite, stannite, PMCA, and three-phase-mixed structures (solid lines), together with $\alpha_{Ex}$ (open circles) and (b) variation of $E_{DFT}$ with stannite volume fraction in CZTSe and CZTS. In (a), the experimental data are consistent with Fig. 3. In (b), the $E_U$ positions are indicated by the dotted lines.

To our knowledge, this is the first result in which the semiconductor tail state is analyzed by applying DFT. Importantly, there has been no clear evidence that the large tail-state absorption in CZTS is caused by cation mixing. As confirmed from Fig. 8(a), our DFT-based analysis provides the first semi-empirical result that the extensive cation substitution in CZTS generates the quite strong tail absorption in a wide energy range near $E_g$. More complete descriptions for the cation disordering in CZTS can be found in Sec. IV B.

Figure 8(b) shows the variation of $E_{DFT}$ with stannite volume fraction in CZTSe and CZTS. For this calculation, the kesterite-stannite two phase composite is assumed. In the figure, the experimental $E_U$ values are also indicated by the dotted lines. Rather surprisingly, $E_{DFT}$ shows a rapid increase up to the stannite fraction of 20 vol.% at which $E_{DFT}$ of the CZTSe becomes comparable to $E_U$. In the case of CZTS, experimental $E_U$ cannot be reproduced by the two phase mixture and thus the three phase crystal structure was assumed. The result of Fig. 8(b) indicates that the slight cation disordering induces a quite large increase of $E_U$ in CZT(S)Se and the quite strong tail absorption can quantitatively be explained by the band gap fluctuation generated by the extensive cation mixing.

**IV. DISCUSSION**

**A. Defect formation in CZTS and CZTSe**

So far, a variety of point and complex defects have been proposed to be generated within CZTS(Se) crystals [30-37]. Among those, a defect created by Cu-Zn cation exchange (i.e., $Cu_{Zn}+Zn_{Cu}$) shows the lowest formation energy for the majority of the chemical potential range [32]. In fact, there has been a strong consensus that $Cu_{Zn}+Zn_{Cu}$ is the most likely defect in CZTS(Se) and the $Cu_{Zn}+Zn_{Cu}$ formation is more energetically favorable, compared with point defects of $Cu_{Zn}$ and $Zn_{Cu}$ [32,37]. It should be emphasized that the tail absorption observed in CZTS(Se) is quite strong with the $\alpha$ values of $10^3 \sim 10^4$ cm$^{-1}$ (see Fig. 3) and the low-concentration defects are very unlikely to induce the strong tail absorption confirmed in CZTS(Se).

At some chemical potential conditions, however, Cu$_{Zn}$ antisite defects (i.e., Zn site replaced with Cu) form more easily [32]. Nevertheless, the formation energy of Cu$_{Zn}$ increases drastically under the Cu-poor and Zn-rich conditions [32], which have typically been employed for the fabrication of high-efficiency CZTS and CZTSe solar cells [31,32]. For a CZTS layer fabricated under the Cu-poor and Zn-rich condition, a quite high $E_U$ of 85 meV, comparable to those of the CZTSe and CZTS layers having near stoichiometric compositions (Fig. 6), has been confirmed [25]. Accordingly, the compositional modulation does not appear to alter $E_U$ and it is unlikely that Cu$_{Zn}$ defects induce the strong tail absorption in CZTS(Se).

On the other hand, the presence of a Cu-vacancy defect complex (V$_{Cu}$+Zn$_{Cu}$) has been confirmed experimentally [34]. However, the DFT calculations show that the formation of this defect complex leads to a slight increase of the effective band gap [32,33,35] and the generation of the tail-state absorption by these defects is not plausible. In addition, the V$_{Cu}$ states are created very close to the valence band (≤20 meV from the valence band) [30,32] and we also ruled out the possibility that the extensive tail absorption is generated by V$_{Cu}$ defects.

In CZTS(Se), the formation of 2Cu$_{Zn}$+Sn$_{Zn}$ defect complex has also been proposed [31]. Nevertheless, this is a deep-level defect and the population range of the defect is predicted to be $10^{11} \sim 10^{18}$ cm$^{-3}$ [32], which is apparently too small to induce the strong tail absorption. Systematic DFT studies have shown that the formation energies of other defects are quite high [32] and the densities of these defects are expected to be low. Based on the above considerations, the exceptionally large absorption observed below $E_g$ of CZTS and CZTSe has been attributed to Cu$_{Zn}$+Zn$_{Cu}$ defects in this study.

**B. Cation disorder in CZTS and CZTSe**

Quite early DFT studies implemented on CZTS and CZTSe confirmed that the kesterite is the most probable crystal structure for CZTS(Se) [18,29]. Nevertheless, the total energies of the kesterite, stannite and PMCA phases are quite similar and the formation of the stannite and PMCA phases by cation intermixing was expected to occur [18,29]. Later, however, such pictures are denied almost completely by detailed structural characterization of experimental CZTS(Se) crystals based on neutron diffraction [38], resonant x-ray diffraction [39], scanning transmission electron microscopy [40], nuclear magnetic resonance [41] experiments and only kesterite-based crystals are found to be formed, although slight cation mixing do occur in the Cu-Zn atomic planes of the kesterite [i.e., the cationic plane position of z=1/4 and 3/4 along the c axis in Fig. 2(a)] [38-41].

However, we emphasize that all the results that conclude the kesterite phase formation without the inclusion of the stannite and PMCA phases have been obtained from the single crystals formed at high temperatures (~800 $^oC$) with sufficient time. In coevaporated CZTS(Se) layers formed at lower temperatures, the cation disordering can be very different from those observed in the single crystals. In order to further support our conclusion that the extraordinary tail absorption observed in coevaporated CZTS and CZTSe layers originates from quite extensive cation substitutions, we have performed DFT calculations for various cation disordered phases shown in Fig. 9.

With the crystals of Fig. 2, the crystal structures shown in Fig. 9 (a)~(d) complete the different Cu-Zn cation disordered (CD) structures in a 16-atom cell. The crystal structure of Fig. 9(a) can be considered as the kesterite phase with a disordered Cu-Zn plane at a z=3/4 position. In the cation disordered structures of Fig. 9(b)~(d), Cu and Zn planes (CD2) and unique Zn clustered structures (CD3, CD4) are created. We also assumed a cation disordered structure that is obtained by exchanging one Cu atom with one Zn atom in a 32-atom kesterite supercell structure [CD 5 of Fig. 9(e)], as suggested from the structural studies of the single crystals.

Table II summarizes the lattice parameter, crystal distortion ($\eta=c/2a$), $E_g$ and total energy difference ($\Delta E_t$) of all the CZTS crystals shown in Figs. 2 and 9. All the CZTS results were deduced based on the PBE calculation using a 8 × 8 × 4 k mesh for a 16-atom-cell and a 4 × 8 × 4 mesh for a 32-atom cell. In the table, the underestimated $E_g$ values in the PBE calculations were corrected by adding 0.40 eV as estimated in the analysis of Fig. 8(a), whereas $\Delta E_t$ values were obtained as a difference from the kesterite structure (i.e., $\Delta E_t=0$ in the kesterite). Figure 10 further shows the $\alpha$ spectra of all the CZTS crystals.

As mentioned above, in the single crystals, the cation mixing occurs only in the limited planes (i.e., at z=1/4 and 3/4). Specifically, in the case of the 16-atom cell, only the kesterite and CD1 structures are allowed and $\Delta E_t$ of CD1 is almost identical to that of the kesterite. However, CD1 indicates $E_g$ and $\alpha$ very similar to those of the kesterite and thus the incorporation of the CD1 phase into the kesterite host crystal does not generate the large tail absorption confirmed in the experiment. When the Zn-plane and Zn-clustered structures are formed (i.e., CD2~CD4 in Table II), on the other hand, $E_g$ shifts significantly to lower energies with the red shift of the whole $\alpha$ spectra.

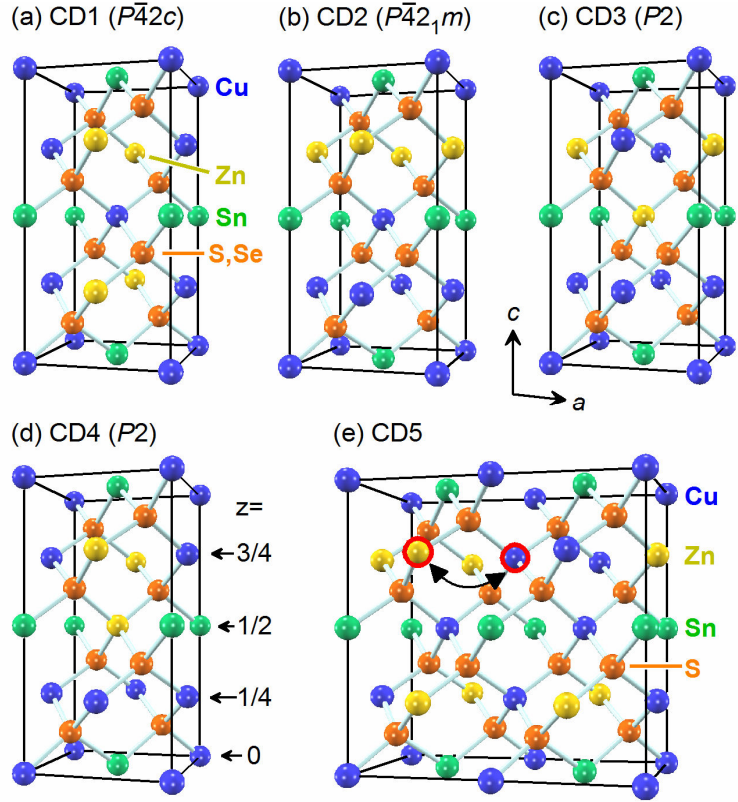

FIG. 9. Cation disordered (CD) structures of CZTS for a 16-atom cell. In a 32-atom supercell shown in (e), one pair of Cu and Zn atoms indicated by the red circles is replaced assuming a kesterite structure.

TABLE II. Lattice parameters ($a$, $b$, $c$), crystal distortion ($\eta=c/2a$), $E_g$ and total energy difference ($\Delta E_t$) of different CZTS crystal structures shown in Figs. 2 and 9.

| Structure | $a$ (Å) | $b$ (Å) | $c$ (Å) | $\eta$ (c/2a) | $E_g$ (eV) [a] | $\Delta E_t$ (meV/atom) |
|---|---|---|---|---|---|---|
| Kesterite | 5.552 | 5.552 | 10.982 | 0.989 | 1.281 | 0.0 |
| Stannite | 5.548 | 5.548 | 11.010 | 0.992 | 1.153 | 3.6 |
| PMCA | 5.530 | 5.530 | 11.054 | 0.999 | 0.972 | 4.5 |
| CD1 | 5.553 | 5.553 | 10.998 | 0.990 | 1.266 | 0.5 |
| CD2 | 5.557 | 5.557 | 11.091 | 0.998 | 1.063 | 24.3 |
| CD3 | 5.556 | 5.554 | 11.010 | 0.991 | 0.921 | 16.9 |
| CD4 | 5.555 | 5.556 | 11.013 | 0.991 | 0.915 | 16.9 |
| CD5 | 11.104 | 5.554 | 10.989 | 0.990 [b] | 1.217 | 4.5 |

a) Calculated by PBE assuming a $E_g$ correction of 0.40 eV, b) Calculated from $c/a$.

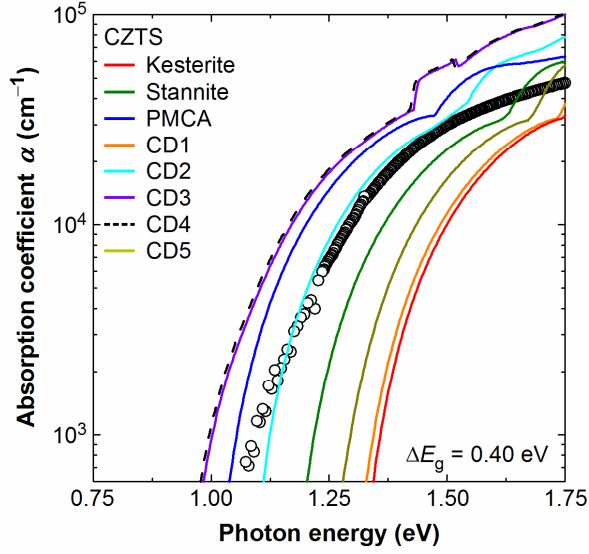

FIG. 10. $\alpha$ spectra calculated for different CZTS crystal structures shown in Figs. 2 and 9 (solid lines) and $\alpha$ spectrum obtained from experiment (open circles).

Nevertheless, these crystals show remarkable increase in $\Delta E_t$ and therefore it is quite unlikely that these phases are formed in experimental crystals. In particular, when the Zn atomic plane is created (CD2), $\Delta E_t$ shows a very high value.

When one pair of the Cu and Zn atoms is replaced in the 32-atom structure (CD5), $E_g$ shifts only slightly by 64 meV toward lower energy, compared with the kesterite. In this structure, $\Delta E_t$ also increases due to the local clustering of Cu and Zn atoms. It should be emphasized that, when the cation mixing at z=1/4 and 3/4 is assumed in a 32-atom spercell, only the pure kesterite [i.e., Fig. 2(a)], CD1 and CD5 structures are allowed. In these structures, there is only a weak $E_g$ shift and thus the cation substitution within the limited cationic planes cannot explain the large tail absorption observed for CZTS(Se).

In earlier studies, the link between the strong tail absorption and cation disordering in CZTS(Se) has been overlooked and our result indicates that the large tail state absorption confirmed experimentally in coevaporated CZTS(Se) layers can best be explained by the mixed-phase formation of the kesterite/stannite/PMCA crystal structures that show the small $\Delta E_t$ values among the possible cation-disordered phases. Although the tail state absorption could still be generated by the incorporation of other cation-mixed phases, $\Delta E_t$ values of such phases tend to be high [42] and the generation of such phase becomes energetically more difficult.

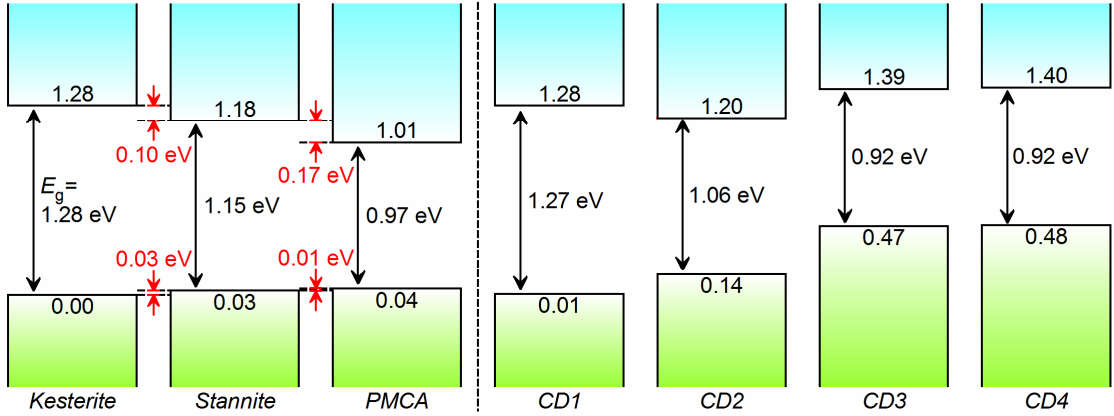

FIG. 11. Band alignments for different CZTS crystal structures shown in Figs. 2 and 9. In the figure, the conduction band positions have been shifted upward by $\Delta E_g$=0.40 eV to compensate the underestimated $E_g$ in the PBE calculations. The numerical values shown within each band represent the relative energy positions of each band.

### C. Potential fluctuation in CZTS

To find the effect of the tail state formation on the solar cell performance more clearly, we calculated the band alignment of the different CZTS crystal structures shown in Figs. 2 and 9. In our calculations, the core level corrections were made simply using the 4$d$ level of the Sn atom [43]. Figure 11 summarizes the band alignments of the CZTS crystals. In this result, the conduction band positions have been shifted upward by $\Delta E_g$=0.40 eV to compensate the underestimated $E_g$ in the PBE calculations.

It can be seen from Fig. 11 that the incorporation of the stannite and PMCA phases into the kesterite leads to the formation of the localized states just below the conduction band minimum as reported for $Cu_2ZnGeS_4$ [19]. A similar result has also been confirmed for the kesterite and stannite crystals of CZTSe. Based on the above result, we conclude that the extraordinary large $E_U$ observed in the CZTS and CZTSe originates primarily from the extensive tail states formed near the conduction band edge.

On the other hand, the conduction and valence band positions of the CD1 structure are almost identical to those of the kesterite structure. Furthermore, in the cation-disordered phases of CD2–CD4, the $E_g$ reduction occurs by the upward shift of the valence band positions and this trend is quite different from that of the stannite and PMCA structures. Due to high $\Delta E_t$ of the CD2−CD4 structures, however, we ruled out the formation of these phases.

In CZTS, a conduction band offset created in a kesterite/stannite/PMCA mixture is quite large with $\Delta E_c$=0.27 eV (see Fig. 11). In other words, the tail absorption in a CZTS crystal extends in a wide energy region of ~0.3 eV below $E_g$, as confirmed directly from the experimental result of Fig. 3. Unfortunately, conventional CZTS(Se) solar cells are fabricated using p-type CZTS(Se) absorbers and thus the minority carriers in the solar cells are electrons. In this case, the tail states formed near the conduction band are expected to act as trap sites for electrons and are quite detrimental in the operation of the solar cells.

**D. $V_{oc}$ deficit in CZTS solar cells**

In CZTS, a quite large $V_{loss}$ of ~0.8 eV has been reported [37,44]. In conventional solar cells including GaAs and Cu(In,Ga)Se$_2$, $V_{loss}$ is in a range of 0.3~0.4 V (see Sec. I) and $V_{loss}$ of CZTS is higher by 0.4~0.5 V. We attributed this additional $V_{oc}$ deficit observed in CZTS solar cells primarily to the band gap fluctuation within the conduction band ($\Delta E_c$ ~ 0.3 eV) due to extensive cation disordering.

Unfortunately, the interpretation of the significant $V_{loss}$ in CZTS(Se) has been controversial and the following has been suggested as the causes of $V_{loss}$ (or performance loss): i) the electrostatic potential fluctuation induced by charge states [1,33,37,45-48], ii) the Cu-Zn cation substitution [42], iii) the formation of deep defects (2Cu$_{Zn}$+Sn$_{Zn}$) [31,36] and iv) the defect formation at the CdS/CZTS(Se) front interface [49,50]. So far, many earlier studies proposed the presence of a strong electrostatic potential fluctuation, characterized by the parallel shift of a constant band gap due to spatial variation of charged defects [1,33,37,45-48]. Such conclusions have been drawn mainly from the detailed analysis of photoluminescence (PL) spectra [45-48]. Nevertheless, PL characterizes the light emission, rather than the light absorption, and the anomalous tail absorption in CZTS(Se) has not been considered in these studies.

The band gap fluctuation by cation mixing was also suggested previously [42]. However, in a recent experimental study that modulated the cation disorder in CZTSSe devices by changing the cooling rate after lower temperature annealing, the effect of the cation disorder on $V_{oc}$ (or band gap fluctuation) has been concluded to be very small (~40 mV) and the cation mixing as a cause of the large $V_{loss}$ has been denied [37]. In the study, however, the influence of the quite strong band tailing has been neglected almost completely.

At this stage, only limited analyses have been performed to clarify the effect of the defect formation on $V_{loss}$ [50] and more quantitative studies are necessary to determine its contribution on $V_{oc}$. As evidenced in this study, however, the energy range of the extensive tail formation (~0.3 eV below the fundamental band gap) can explain very large $V_{loss}$ confirmed in CZTS devices. Accordingly, the suppression of the strong cation disordering is expected to be crucial for the further improvement of CZTS(Se) solar cells.

## V. SUMMARY

The absorption edge energies of various solar-cell absorber materials, including CdTe, CISe, CGSe, CZTS, CZTSe and hybrid perovskite compounds, have been calculated by DFT to determine the tail state formation in the absorber materials. Very high-density k meshes have been used in these DFT calculations to characterize the absorption edge accurately. The absorption edge energy deduced from the DFT calculations indicates an excellent correlation with the Urbach energy, confirming the formation of ideal sharp band edges in experimental crystals. In particular, MAPbI$_3$ hybrid perovskite shows the sharpest absorption edge theoretically, indicating superior band-edge transition properties of this material. In contrast, we observe that the Urbach energies of polycrystalline CZTSe and CZTS are far larger than the theoretical values. The very large Urbach energies observed experimentally in CZTS(Se) have been attributed to the cation substitution which in turn generates the tail state formation near the conduction band edge. In particular, for the CZTS, by taking a weighting average of three DFT $\alpha$ spectra obtained for kesterite, stannite and PMCA phases, the experimental $\alpha$ spectrum has been reproduced. As a result, our theoretical approach is found to be quite effective in determining band tailing originating from imperfect crystal formation.

*Supplementary Materials*

**Tail state formation in solar cell materials: First principles analyses of zincblende, chalcopyrite, kesterite and hybrid perovskite crystals**

Mitsutoshi Nishiwaki,[1] Keisuke Nagaya,[1] Masato Kato,[1] Shohei Fujimoto,[1] Hitoshi Tampo,[2] Tetsuhiko Miyadera,[2] Masayuki Chikamatsu,[2] Hajime Shibata,[2] and Hiroyuki Fujiwara[1*]

[1]*Department of Electrical, Electronic and Computer Engineering, Gifu University, 1-1 Yanagido, Gifu 501-1193, Japan*

[2]*Research Center for Photovoltaics, National Institute of Advanced Industrial Science and Technology (AIST), Central2, 1-1-1 Umezono, Tsukuba, Ibaraki 305-8568, Japan*

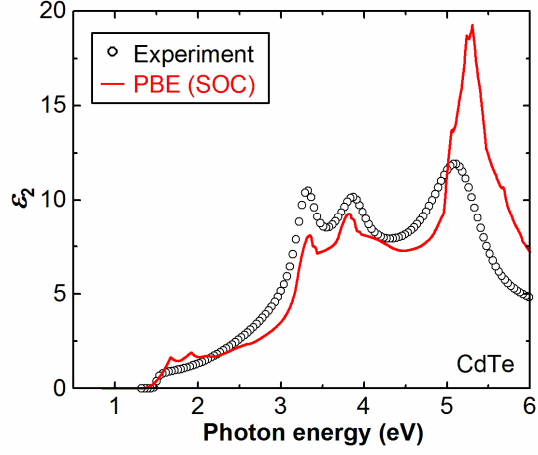

**FIG. S1.** $\varepsilon_2$ spectrum calculated by PBE with spin-orbit coupling (SOC) interaction (solid line). The open circles indicate the experimental data. The PBE-SOC result has been blue shifted by 0.85 eV.

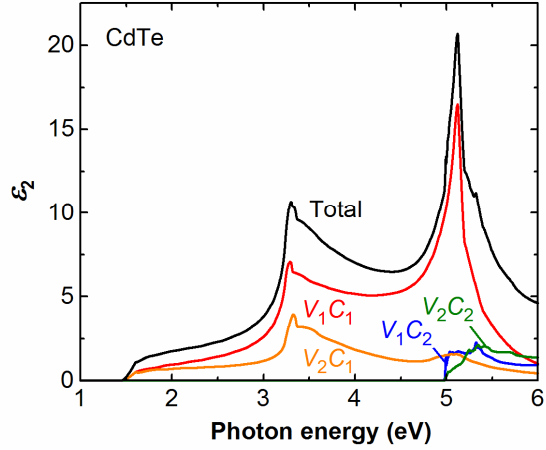

**FIG. S2.** Optical transition analysis for CdTe. The $\varepsilon_2$ contributions for each transition from the valence band to the conduction band are shown. The $V_jC_k$ in the figure denotes the transition from the $j$th valence band to the $k$th conduction band. The $\varepsilon_2$ spectrum (total) is consistent with that shown in Fig. 4(b).

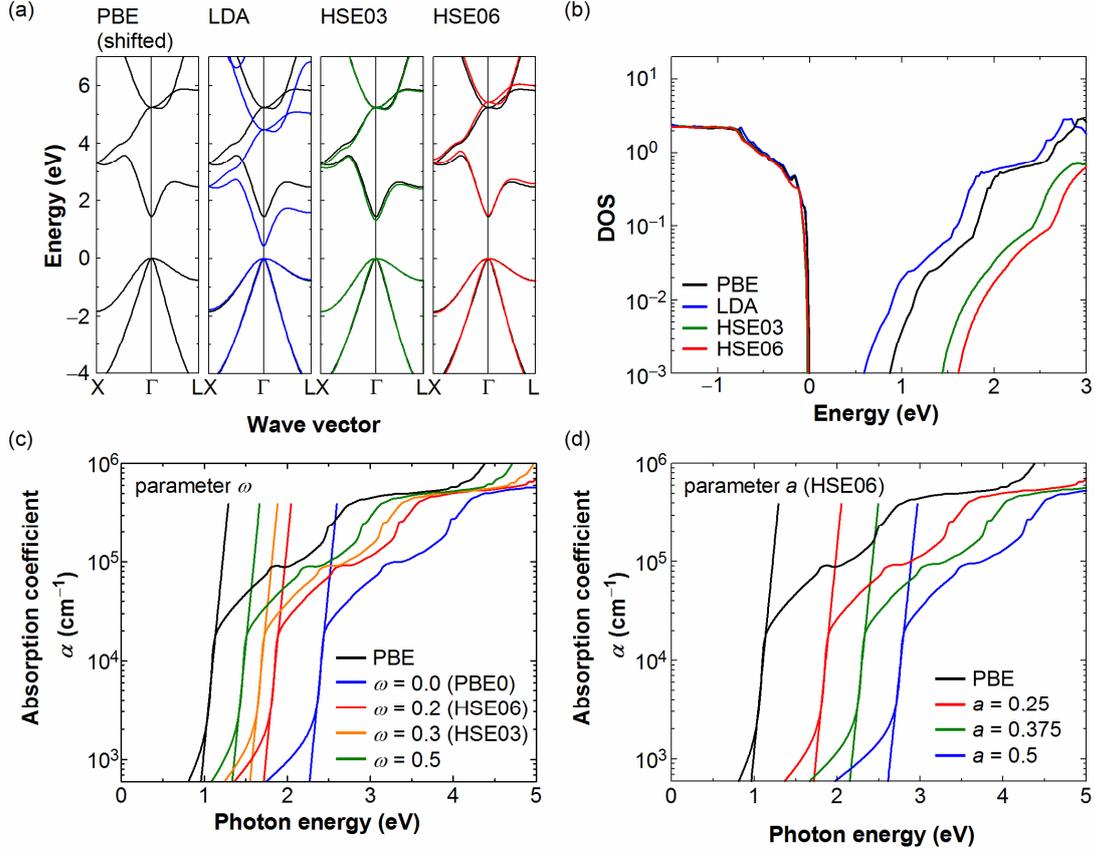

**FIG. S3.** (a) Band structures, (b) DOS, (c) $\alpha$ with the variation of the screening parameter ($\omega$) of the hybrid functional and (d) $\alpha$ with the variation of the mixing parameter ($a$) of HSE06, obtained for CdTe. In (a), the black lines show the shifted PBE result, which is consistent with Fig. 4(a). In (c), $\omega$ was changed in a range of 0~0.5 Å$^{-1}$, whereas the parameter $a$ for HSE06 was varied from 0.25 to 0.5 in (d). The results of (a) and (b) were obtained using a 10 × 10 × 10 k mesh, whereas a 8 × 8 × 8 k mesh was employed for the calculations of (c) and (d) due to a high computational cost. The solid lines in (c) and (d) indicate the analysis results for $E_{DFT}$. In (c), the $E_{DFT}$ values are quite constant (50 meV) with a difference less than 1 meV. The $E_{DFT}$ is also constant with the variation of $a$ ($E_{DFT}$=51−55 meV). It should be noted that the small absorption tail observed at $\alpha$<3×10$^3$ cm$^{-1}$ is artifact and is generated by the spectral smoothening incorporated in the VASP code. We minimized this error by reducing the smoothening parameter value (i.e., by decreasing the CSHIFT parameter to 0.02).

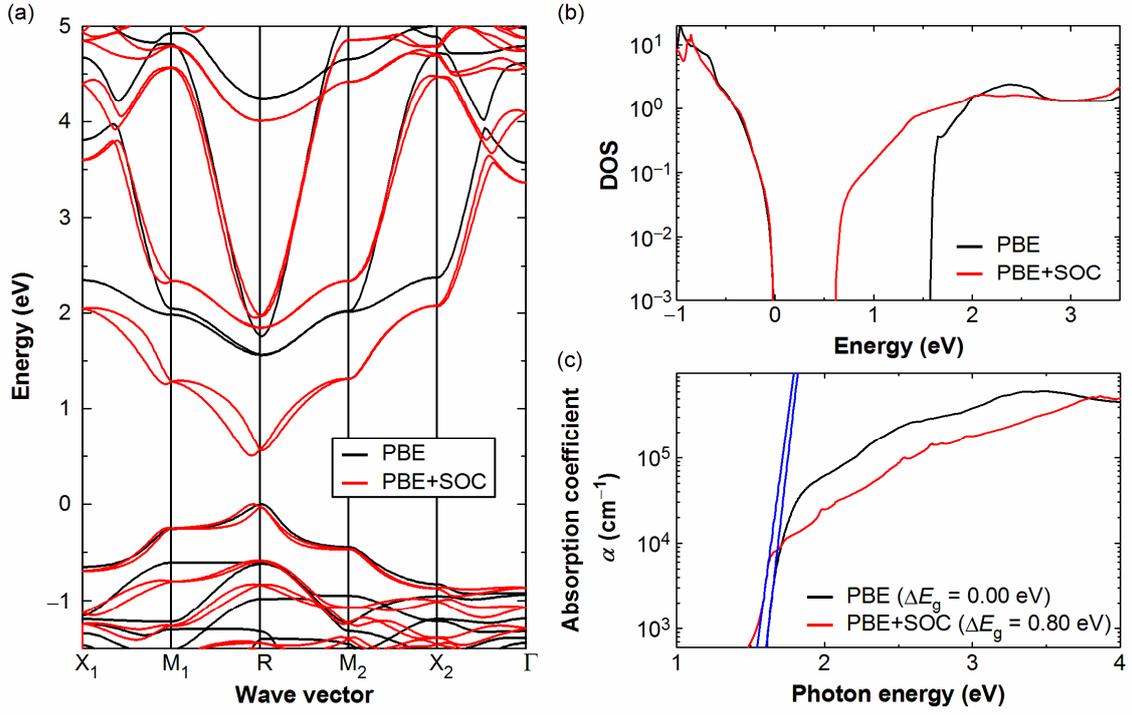

**FIG. S4.** (a) Band structures, (b) DOS and (c) $\alpha$ spectra of MAPbI$_3$ calculated within PBE with and without SOC. In (c), the DFT calculations were implemented using the same 8 × 8 × 8 k mesh and the solid lines indicate the result of the $E_{\mathrm{DFT}}$ analyses, which yield similar $E_{\mathrm{DFT}}$ values of 29 meV (PBE) and 33 meV (PBE+SOC).